\documentstyle[preprint,aps]{revtex}
\begin{document}
\draft
\title{
Mode-matching technique for transmission calculations in
electron waveguides at high magnetic fields}
\author{J. J. Palacios and C. Tejedor}
\address{
Departamento de F\'{\i}sica de la Materia Condensada.
Universidad Aut\'onoma de Madrid.
Cantoblanco, 28049, Madrid. Spain.}
\maketitle
\begin{abstract}
In this paper we present a mode-matching technique
to study the transmission
coefficient of mesoscopic devices such as electron waveguides in the
presence of high magnetic fields for different situations.
A detailed study of the difficulties rising due to the presence of the
magnetic field is given and the differences with the zero magnetic field
case are stressed.
We apply this technique to calculate the transmission
 at non-zero magnetic
field of two completely different systems: a) a quantum box built up on
a quantum wire (or electron waveguide) by means of two
barriers and b) a meandering quantum wire, {\em i.e.}, a
wire with changes in the guiding direction. In the former case we analyze
the so-called {\em Coulomb Blockade} and
{\em Aharanov-Bohm} regimes and in the
latter one we focus on the realistic case of soft, circular
bends joining the different sections of the wire.
\end{abstract}
\pacs{73.40.Cg, 73.50.Jt}

\section{Introduction}
\label{intro}

Landauer-B\"{u}ttiker formula relates, in a simple manner,
the linear response
conductance with the transmission coefficient of an electronic system
and it has been applied successfully in a number of transport problems in
mesoscopic devices \cite{ssp441}. Only two ingredients are needed:
a) reservoirs
in which thermalization may occur and b) a region, whose transport
properties
we want to know, free of inelastic scattering. The transmission
coefficient
of this region is directly responsible for the conductance in such a way
that elastic processes are the only ones to be taken into account.

In order to calculate the
transmission, many methods have been developed over the past few years.
On the one hand, tight-binding-like techniques \cite{zpb59385},
in which space is discretized into cells, have been applied
satisfactorily
in many problems, especially, in those including
disorder \cite{prb448017}. On the other hand,
mode-matching methods have been used in problems
like those of disorder-free electron
waveguides. Conductance calculations for a ballistic constriction in the
absence \cite{ssc68715} and in the presence of a
magnetic field \cite{prb403429,prb4513725}
can be done
with this method. Following with the so-called two-probes systems
as the one just mentioned, Bagwell dealt with the effects on the
conductance of a quantum wire of one and two $\delta$-function-like
scattering centers \cite{prb4110354,prb439012}.
Changes in the confining geometry of a quantum
wire is also a suitable problem to be treated within the framework of
the matching technique; for instance, cavities studied by Kasai {\em et al.}\
\cite{jpsj601679} and Wu {\em et al.}\ \cite{prb446315}
and single and multiple bends \cite{prb4111887,jap71515,prb4511960,prevko}.
One-dimensional periodic structures have been also studied
with this technique \cite{prb4312082}.
As for cross wires or junctions, encountered in three and four-probes
devices, this method has revealed itself extraordinarily useful
to understand phenomena such as negative resistances \cite{prl622527},
resonant tunneling through bound
states in open systems \cite{prb395476}, quenching of the Quantum Hall
Effect (QHE) \cite{prb4112760},
box resonators in crossed wires \cite{prb448399}, etc.

Most of the work above mentioned has been done at zero magnetic field.
This fact reduces the interest of it since many interesting phenomena
in mesoscopic physics derive from the QHE \cite{prange} and
are related with the possibility of applying a magnetic field
perpendicular to the two-dimensional electron gas (2DEG).
The difficulties inherent to the magnetic field can be almost completely
removed if waveguides with parabolic confinement potential are
considered for the calculations \cite{jpsj601679,jpsj592884}.
However, screening properties due
to the electron-electron interaction \cite{sslfs} show that, at not a
very low density, the shape of the effective confinement potential
is rather
flat in the middle of the quantum wire and rises quickly at
the edges of it (independently of the fact of having been defined
either by split-gates \cite{jpcm16291},
or by etching \cite{thkou}). Additional step-like structure
may appear in the
classically calculated \cite{prb464026}, or
self-consistently calculated confining potential \cite{prebpt} but we
will not consider this possibility in this paper in order to simplify our
model.
Bearing this fact in mind we
consider appropriate to simulate the confinement by means of a square
hard wall
potential as done in Refs.\ \cite{prb4112760,prevko}.

Sec.\ \ref{match} is devoted to the discussion of
the numerical details for this case. In Sec.\ \ref{box} we analyze the
conductance properties of a quantum box in the presence of a magnetic
field and in Sec.\ \ref{bends} those of a twisting quantum wire
with circular, soft bends. Finally, the conclusions of our work
are presented in Sec.\ \ref{conclusions}.

\section{MATCHING TECHNIQUE IN MAGNETIC FIELDS}
\label{match}

The starting point of a matching method relies on obtaining the
most general wave function
for all the different regions involved in the problem according to
an usual scattering-like problem.
We consider a waveguide in such a way that electrons move freely in the
$y$ direction and feel hard walls in the $x$ one. In Fig.\ \ref{fig1}
such a waveguide is shown including a region of arbitrary potential
in the middle of it. The Schr\"odinger equation in the Landau gauge
\begin{equation}
\mbox{{\boldmath $A$}} = (0,Bx\;l_m,0)
\end{equation}
becomes separable in the $x$ and $y$ variables for
each different region $I,II,III$, being characterized
respectively by different confinement potential profiles
$V^{I,II,III}(x)$, and it looks like:
\begin{equation}
-\frac{\partial^{2}\phi^{I,II,III}_{n}(x)}{\partial x^{2}}
+ (k^{I,II,III}_{n} + x)^{2}\phi^{I,II,III}_{n}(x) +
v^{I,II,III}(x)\phi^{I,II,III}_{n}(x) = \epsilon\phi^{I,II,III}_{n}(x)
\label{schr}
\end{equation}
with
\begin{equation}
\Phi^{I,II,III}_{n}(x,y) = e^{ik^{I,II,III}_{n}y}\phi^{I,II,III}_{n}(x)
\label{func1}
\end{equation}
being the total wave function for a given wave vector $k_{n}$ in each region
$I,II,III$ where $x$ and $y$ are given in units of the magnetic length
$l_{m}=\sqrt{\hbar/eB}$, wave vectors $k_{n}$ in $l_{m}^{-1}$,
$\epsilon =
2E/\hbar\omega_{c}$,
$v^{I,II,III}(x)=2V^{I,II,III}(x)/\hbar\omega_{c}$
($\omega_{c}$ is the cyclotron frecuency $eB/m^{*}$ for particles
of charge
$e$ with effective mass $m^{*}$ under a magnetic field $B$ and
$E$ is the energy). Spin splitting effects due to
the magnetic field are neglected for the GaAs-AlGaAs  heterostructure
considered bellow in all the calculations.

In the case depicted in Fig.\ \ref{fig1} both regions I and III
are identical
and the potential profile in the $x$ direction
for both of them is depicted in Fig.\ \ref{fig2}a.
The dispersion relations for real wave vectors for the
lowest subbands have been depicted too. The
energy was fixed in such a way that there are three subbands
occupied or three
current carrying channels. Now the problem consists of finding the
wave vectors $k_{n}$ (we drop labels $I,II,III$) and corresponding
transversal modes $\phi_{n}(x)$ for a given energy. Finding the
complex band structure is a simple
problem in the case of zero magnetic field (for any confinement
potential in the $x$ direction) since
wave vectors and modes can be found
analytically thanks to the usual parabolic dispersion
relations (free electrons in the $y$ direction).
It is also simple the case of
a parabolic confinement potential in the presence of a  magnetic
field \cite{jpsj601679,jpsj592884}. However,
if the dispersion relations are no longer parabolic, the problem must be
solved numerically as it was already stressed in Ref.\ \cite{prb4112760}.
Wave vectors $k_{n}$ are the solutions of the intersection between the
complex dispersion relations (real part of it is shown in Fig.\
\ref{fig2}a)
and the plane of constant energy. Details on this calculation
are given in the Appendix and the result of this intersection is shown in
Fig.\ \ref{fig2}b. Solid dots denote the wave vectors
in the $k$-complex plane.
Those lying on the real axis (with null imaginary part) correspond
to extended modes and those with non-zero imaginary part
correspond to evanescent or exploding modes.
The latter ones can be grouped into two
types depending on whether its real part is null or not. The
former ones belong to the subbands (four for each subband) with a dispersion
relation having
flat regions (bulk regions within the terminology of the QHE)
and the latter
ones belong to the rest of subbands
with a minimum in their dispersion relation (two for each subband
\cite{aclarar}).
This picture is modified as we change the magnetic field
in the following way:
If we increase the magnetic field dots with null real part split into two
(arrows pointing outwards
showing this fact) and, if we decrease it, double dots collapse
into one (arrows pointing inwards). Turning off the field completely, purely
real wave vectors and purely imaginary ones appear in the $k$-complex
plane corresponding to parabolic dispersion relations of subbands below or
above the chosen energy respectively. The total number of subbands
$N$ is given by the sum of $n_{1}$ (number of
subbands below the chosen energy), $n_{2}$ (number of
subbands above that energy with flat regions in their
dispersion relation) and $n_{3}$
(the rest of them having a minimum in their dispersion
relation). So, $n_{T}=2n_{1}+4n_{2}+2n_{3}$ is the total number of
wave vectors involved in the $N$-subbands problem. Sometimes, this fact
has not been duly appreciated in tight-binding-like
schemes \cite{prb448017}.

It is also shown in the Appendix how the $\phi_{n}(x)$ functions
(transversal modes) can
be found once we know the $k_{n}$ wave vectors. The above discussion has
focused on regions $I$ and $III$ but is valid in general
terms for region $II$ provided that its potential is only dependent
on $x$ and not on $y$, in order for the Schr\"odinger equation to be
separable.
In this way, the most general wave function at a given energy
can be expressed like:
\begin{equation}
\left. \begin{array}{ll}
\mbox{Region $I$:} & \Phi^{I}(x,y)=\sum_{n=1}^{\infty}\alpha_{n}
\Phi_{n}^{I}(x,y)  \\
\mbox{Region $II$:} & \Phi^{II}(x,y)=\sum_{n=1}^{\infty}\beta_{n}
\Phi_{n}^{II}(x,y) \\
\mbox{Region $III$:} & \Phi^{III}(x,y)=\sum_{n=1}^{\infty}\gamma_{n}
\Phi_{n}^{III}(x,y)
\end{array} \right\}
\label{equ3}
\end{equation}
The coefficients in the above expressions must be
determined by matching on
the delimiting interface but
there are additional conditions inherent to the scattering problem and,
as far as our problem is concerned,
not all the solutions present in the $k$-complex plane can be considered.
In region $I$ the incident and reflecting modes must be taken into
account (those on the left and right arms of the real axis respectively)
so as the non-exploding ones (dots in the lower
part of the $k$-complex plane in Fig.\ \ref{fig2}b).
In region $II$ there is no restriction and all the solutions must be
taken into account and
in region $III$ only out-going modes (dots on the real
axis on the left arm)
and those with dots in the upper half plane must be considered. So,
the solution looks like this:
\begin{equation}
\left. \begin{array}{ll}
\mbox{Region $I$:}&\Phi^{I}(x,y)=\Phi_{i}^{I}(x,y) +
\sum_{j=1}^{\infty}r_{ij}\Phi_{j}^{I}(x,y)  \\
\mbox{Region $II$:}&\Phi^{II}(x,y)=\sum_{j=1}^{\infty}\beta_{ij}
\Phi_{j}^{II}(x,y) \\
\mbox{Region $III$:}&\Phi^{III}(x,y)=\sum_{j=1}^{\infty}t_{ij}
\Phi_{j}^{III}(x,y)
\end{array} \right\}
\end{equation}
where a single incident mode $i$ has been chosen (the same must be done
for all the incident modes) and the surviving coefficients from those
$\alpha_{n}$, $\beta_{n}$ and $\gamma_{n}$
appearing initially in Eq.\ \ref{equ3}
have been relabeled as $r_{ij}$,
$\beta_{ij}$ and $t_{ij}$.

Now the matching consists of the standard problem of invoking continuity
of the wave function and its derivative across the two interfaces delimiting
the three regions:
\begin{eqnarray}
\Phi_{i}^{I}(x,0) + \sum_{j=1}^{\infty}r_{ij}\Phi_{j}^{I}(x,0) &
= & \sum_{j=1}^{\infty}\beta_{ij}\Phi_{j}^{II}(x,0) \\
\sum_{j=1}^{\infty}\beta_{ij}\Phi_{j}^{II}(x,L) &
= & \sum_{j=1}^{\infty}t_{ij}\Phi_{j}^{III}(x,L) \\
\left| \frac{\partial}{\partial y}
\left\{ \Phi_{i}^{I}(x,y) + \sum_{j=1}^{\infty}r_{ij}\Phi_{j}^{I}(x,y)
\right\} \right|_{y=0} & = & \left| \frac{\partial}{\partial y}
\left\{ \sum_{j=1}^{\infty}\beta_{ij}\Phi_{j}^{II}(x,y)\right\}
\right| _{y=0} \\
\left| \frac{\partial}{\partial y}
\left\{ \sum_{j=1}^{\infty}\beta_{ij}\Phi_{j}^{II}(x,y)\right\}
\right| _{y=L} & = & \left|\frac{\partial}{\partial y}
\left\{ \sum_{j=1}^{\infty}t_{ij}\Phi_{j}^{III}(x,y)\right\} \right| _{y=L}
\end{eqnarray}

By projecting the obtained equations onto a basis of
states in the $|\psi_{m}\rangle$
(for instance those of the problem without magnetic field, {\em i.e.}
sines and cosines in the $x$ direction)
we reduce the problem to that of solving a
non-homogeneous system of equations
(by picking as many $|\psi_m\rangle$ states
as necessary) from which we obtain the transmission coefficients:
\begin{eqnarray}
\langle \psi_{m}|\phi_{i}^{I}\rangle & = &
- \sum_{j=1}^{\infty}r_{ij}\langle \psi_m | \phi_j^I \rangle +
\sum_{j=1}^{\infty}\beta_{ij}\langle \psi_{m}| \phi_{j}^{II}\rangle \\
0&=&-\sum_{j=1}^{\infty}\beta_{ij}
e^{ik_{j}^{II}L}\langle \psi_{m}|\phi_{j}^{II}\rangle
+ \sum_{j=1}^{\infty}t_{ij}e^{ik_{j}^{III}L}\langle
\psi_{m}|\phi_{j}^{III}\rangle \\
ik_{i}^{I}\langle \psi_{m}|\phi_{i}^{I}\rangle & = & -
\sum_{j=1}^{\infty}ik_{j}^{I}r_{ij}\langle \psi_{m}|\phi_{j}^{I}\rangle
+  \sum_{j=1}^{\infty}ik_{m}^{II}\beta_{ij}
\langle \psi_{m}|\phi_{j}^{II}\rangle \\
0 & = & -\sum_{j=1}^{\infty}ik_{j}^{II}e^{ik_{j}^{II}L}
\beta_{ij} \langle \psi_{m}|\phi_{j}^{II} \rangle
+  \sum_{j=1}^{\infty}ik_{m}^{III}e^{ik_{j}^{III}L}t_{ij}
\langle \psi_{m}|\phi_{j}^{III}\rangle
\end{eqnarray}

Of course, the matching technique has its limitations and is not adequate
if one has to separate the problem in too many different regions or
regions too large compared with the magnetic length. Numerical errors due
to the exponential behavior of the evanescent modes rise unavoidably in the
linear system of equations to be solved for those situations.

\section{QUANTUM BOXES}
\label{box}

The case in which region $II$ is a simple flat barrier can be solved easily
with the expressions in Sec.\ \ref{match}. In this section,
we extend our scheme to the problem of a quantum box defined by means of
two barriers crossing a wire of width $W$ separated by a distance $D$
(we restrict here to the symmetrical case of equal barriers).
This is shown with the inset in Fig.\ \ref{fig3}a and
it is intended to modelize real systems \cite{jpcm16291,thkou}. Once we
know the scattering
matrices $t \equiv t_{ij}$ and $r \equiv r_{ij}$
for a single barrier it is easy
to obtain the total transmission ($T$) and reflection ($R$) matrices:
\begin{eqnarray}
T&=&t\;p\;(1-w)^{-1}\;t \label{T}\\
R&=&t\;p\;r\;p\;(1-w)^{-1}\;t+r \label{R}
\end{eqnarray}
where $w=r\;p\;r\;p$ and $p$ is the diagonal propagation
matrix whose diagonal elements are $e^{ik_{i}D}$
with $k_{i}$ being the wave vectors of all the transmitted modes, extended
and evanescent, in the region between barriers.
The evanescent ones are important if the barriers are nearby each other
or the energy is below but close to the bottom of any subband as it
was already pointed out for the case of a waveguide with
two point-like scatterers \cite{prb439012}. The subspace of extended
modes in the total transmission matrix at a given Fermi energy $E_F$
gives us directly the linear response conductance of the system
through the Landauer-B\"uttiker formula:
\begin{equation}
G=2e^{2}/h\sum_{i,j}^{n_{1}}\frac{v_{j}}{v_{i}}|T_{ij}(E_F)|^{2}
\label{land}
\end{equation}
with $v_{j}$ and $v_{i}$ being the velocities of the outgoing and incoming
modes respectively. A trivial expression for the velocity of these modes
(in units of $l_ms^{-1}$) can be obtained from the Hellman-Feynman theorem:
\begin{equation}
v_{i}=\omega_{c}(k_{i}- \overline x)
\end{equation}
where $\overline x$ is the center of gravity of the mode.

Figs.\ \ref{fig3}a and \ref{fig3}b show
the conductance as a function of the Fermi energy
for two different values of the magnetic field. The values of $W$ and
$D$ are $4$ and $8\:l_{m}$ (for the case of $1$ T)
respectively, which roughly corresponds to a box
of dimensions $0.1$ and $0.2 \: \mu$m respectively. In
actual experiments the magnetic field is usually swept or,
alternatively, a voltage applied to a bottom or top metal gate. In the
latter cases,
the effect of the sweeping can be simulated by changing the Fermi
energy as we have done in our calculations.
There clearly appear two different regions (hereafter labeled as
region A and region B).
The inset in Fig.\ \ref{fig3}a show the semiclassical or
adiabatic picture of edge channels running along the boundaries.
Although this is not the case in our geometry any more,
it will serve as a visual help for
clarifying purposes. Region A corresponds to the case of having only one
Landau level occupied in the three regions (leads and box) with a kinetic
energy of the edge state not high enough for the electron to pass over the
barrier. That means that any non-zero transmission through
a single barrier will be due to a tunneling process. So, the peaks in the
conductance of the two barriers defining the box correspond to resonant
tunneling through quasi-bound states in the box.
These resonances appear as a consequence of the
term $(1-w)^{-1}$ accounting
for the multiple reflections within the box. It can be equally seen as due
to the poles of the Green's function in the box within the framework of the
Generalized Transfer Hamiltonian (GTH) \cite{prb3810507}.
These states
do not correspond to the semiclassical idea of edge channels unless the
magnetic field is high enough as it can be drawn from Fig.\ \ref{fig4},
in which, the evolution of one of these states is shown as the magnetic
field is increased. At low fields they resemble
to those of a two-dimensional,
field-free square well but increasing the field they evolve to
its edge state-like nature.

However, in this region A, Coulomb effects are extremely important since
the number of electrons confined in the box is a well defined quantity and
corresponds to an integer number (in our case of the order of 10).
This fact gives rise to the so-called Coulomb blockade phenomenon \cite{sct}.
It is not the aim of this paper to discuss quantitatively
these Coulomb effects on the conductance of our system although a few words
can be said on the issue. It has already been discussed
by Palacios {\em et al.} \cite{prept} that the influence of Coulomb
interaction on the conductance of this system (as to the region A
is concerned) is summarized in the following facts:
a) The position of the peaks is
shifted by a charging or activation energy that varies
in a non-trivial way with the number
of electrons in the box and with the magnetic field and it can not be
described in terms of classical capacitances or Anderson models; b) the
height of the peaks is not only reduced from $2$ to $1$ (in units of $e^2/h$)
as classical theories predict \cite{sct,prb441646}
but it is reduced drastically even more due to the strong
correlation suffered by the small
number of electrons in this box. This regime of few electrons is being
reached in recent experiments \cite{thkou,prep} but
a complete study of it is lacked.

Let us now pay attention to the region B in which two Landau levels are
occupied. The first one has enough kinetic energy to run over the barrier
but the second one can not overcome it. Again, the semiclassical picture is
helpful although not completely correct as it will be seen below.
In the cases depicted in both Fig.\ \ref{fig3}a and Fig.\ \ref{fig3}b  there
is an almost flat
region at a fixed value of $2e^{2}/h$ corresponding to the first Landau
running free over the barriers. In the case of $1$ T (Fig.\ \ref{fig3}a) three
peaks appear superimposed to this plateau and they correspond to resonant
tunneling of the edge state belonging to the second Landau level.
This phenomenon has been known as a particular {\em Aharanov-Bohm}
effect in singly-connected systems \cite{prl622523}. They do
not reach the value of $4e^{2}/h$ because of the non-adiabatic behavior
of the edge states in this geometry. There is a mixing
between them which avoids perfect resonance. In Fig.\ \ref{fig3}b
we show the results for a higher value of the magnetic field.
The two peaks at lower $E_F$ on the plateau in Fig.\ \ref{fig3}a
have become dips while the third
one remains as a peak but narrower than before. That can be understood in the
following way: the higher the magnetic field is, the more localized the
states are, so that their coupling to the leads is smaller. This fact
narrows the peaks (as it can be equally
seen in region A). Once these states have become confined enough,
they are used as routes for the first Landau edge states to backscatter, so,
reducing the conductance bellow the value of the plateau. Instead of a
resonant transmission, this phenomenon is a
{\em resonant backscattering}. The peaks on the conductance plateau
are also characteristic of any dot with
an adiabatic potential profile but the dips appearing in Fig.\ \ref{fig3}b
are only due to the non-adiabatic behavior of the edge states
in this geometry and have been
reported in several experimental works \cite{jpcm16291,jpcm13369}.
Recently \cite{prl691989,prb467236},  Coulomb effects have been
observed in this regime although the charge within the dot is not an
integer quantity any more due to the
presence of traveling channels along the dot. These results
have been discussed theoretically \cite{premb} in terms of
compressible and incompressible
regions for the case of a circular geometry with soft walls but this is not
the case presented here.

Figure \ref{gb} shows results of conductance, for the same box,
in the case of
sweeping the magnetic field for a given Fermi energy of
6 $\hbar\omega_c/2$. This is the
most commonly found experimental situation and, from Fig.\ \ref{gb}, we
can see how the results are qualitatively the same as those presented
in Figs.\ \ref{fig3}a and \ref{fig3}b. So, the
above discussion remains valid in general terms. Notice, only, the appearance
of additional
structure in the conductance, for high values of the magnetic field, close
to the region of resonant tunneling. This structure is due to backscattering
of the only carrying-current mode remaining
at those fields and only appear at high
enough fields. As can be seen from Figs.\ \ref{fig3}a and \ref{fig3}b, at
the fields shown there, no structure appear close to the region A of resonant
tunneling.

As to the individual behavior of the transmission coefficients $T_{ij}$,
it suffices to say that they remain diagonal for the whole range of values of
$B$ except for the values at which peaks in the conductance appear.
At those values, the four coefficientes $T_{11}$,$T_{12}$, $T_{21}$ and
$T_{22}$ (for the case of two Landau levels occupied) share equally the
transmission. It is there that the nonadiabatic behavior
of the edge states in this geometry becomes more notorious.

\section{MEANDERING WIRES}
\label{bends}

As another example of the matching technique we have chosen that of a
meandering
wire with soft bends. The case of one bend without magnetic field has
already been
studied by Sols {\em et al.}\ \cite{prb4111887}
and Sprung {\em et al.}\ \cite{jap71515}. We present here the
general way of treating multiple bends in magnetic fields
and we show results for the
simplest case of a single one. In Fig.\ \ref{fig5}a it is illustrated what we
understand for a soft bend. This shape is the adequate one
for the matching and it can be expected to represent that of a real bend in
a waveguide. Unlike the situation presented in Sec.\ \ref{match} this
time the matching is going to be done between two (instead of three)
different regions (Fig.\ \ref{fig5}b). This fact will be justified later on.
There is an additional difference: the interface between regions
does not present a jump in the potential along the propagation direction as
before but presents a change in the confining geometry. The Schr\"odinger
equation in region $I$ is the same as Eq.\ \ref{schr} but in region
$II$, due to the new boundary conditions, we must use polar coordinates
with the symmetrical gauge
\begin{equation}
\mbox{\boldmath {$A$}} = (0,Br\;l_m/2,0)
\end{equation}
in order to separate
the Schr\"odinger equation by using the wave function
\begin{equation}
\Phi_{n}^{II}(r,\theta) = e^{im_{n}\theta}\chi_{n}(r)
\end{equation}
for a given $m_{n}$.  Defining $\chi_{n}(r)=\psi_{n}(r)/\sqrt{r}$ we obtain:
\begin{equation}
-\frac{\partial^{2}\psi_{n}(r)}{\partial r^{2}}
+ \left[-\frac{1}{4r^{2}}+\frac{1}{r^{2}}
\left(\frac{r^{2}}{2}-m_{n}\right)^{2}\right]\psi_{n}(r)+
v(r)\psi_{n}(r) = \epsilon\psi_{n}(r)
\label{schrc}
\end{equation}
The units have been chosen in the same manner as
those of Eq.\ \ref{schr} and the search of the complex $m_{n}$
quantum numbers is undertaken in exactly the same way as that of the
$k_{n}$ wave vectors in Sec.\ \ref{match}. All the facts
concerning the matching in Sec.\ \ref{match} remain valid but now
everything is done for only one delimiting interface:
\begin{eqnarray}
\Phi_{i}^{I}(x,0) + \sum_{j=1}^{\infty}r_{ij}\Phi_{j}^{I}(x,0) &
= & \sum_{j=1}^{\infty}t_{ij}\Phi_{j}^{II}(r,\theta) \\
\left| \frac{\partial}{\partial y}
\left\{ \Phi_{i}^{I}(x,y) + \sum_{j=1}^{\infty}r_{ij}\Phi_{j}^{I}(x,y)
\right\} \right|_{y=0} & = & \left| \frac{1}{r} \frac{\partial}
{\partial \theta} \left\{ e^{ir^{2} \sin \theta \cos \theta/2}
\sum_{j=1}^{\infty}t_{ij}\Phi_{j}^{II}(r,\theta)\right\}
\right| _{\theta=0} \\
\end{eqnarray}
These equations look similar to those in Sec.\ \ref{match}
with the exception of the term
$e^{ir^{2}\sin \theta \cos \theta/2}$ accounting for the required
unification of the
gauge. The matching must be done in a unique gauge, {\em i.e.},
wave function in region $II$ must be expressed in the same gauge
used for region $I$ and vice versa.
Again, we must reduce the expressions above to a non-homogeneous system of
equations by projecting onto a given basis in order to obtain the scattering
matrices $t_{ij}$ and $r_{ij}$. It must be stressed that now the problem
does not present symmetry in the sense that the incident mode can be chosen
in two different ways, any of them giving different resulting scattering
matrices. For instance, the scattering matrices corresponding to incident
modes from region $I$ will be denoted by
$t_{ij}^{iI}$ and $r_{ij}^{iI}$ and those corresponding to incident modes
from region $II$ by
$t_{ij}^{oII}$ and $r_{ij}^{oII}$ (Fig.\ \ref{fig5}b).
The superscripts $i$ and $o$ denote incident
modes along the inner side of the waveguide or along the outer side.
Reversing the magnetic field we obtain a
different set of scattering matrices: $t_{ij}^{oI}$, $r_{ij}^{oI}$,
$t_{ij}^{iII}$ and $r_{ij}^{iII}$.

All these matrices are necessary in order to calculate the transmission or
reflection of the single bend (Fig.\ \ref{fig5}a). To achieve this, Eq.\
\ref{T} and Eq.\ \ref{R} must be generalized in the following way (dropping
the subscripts $ij$):
\begin{eqnarray}
T^{i}&=&t^{iII}p^{i}(1-w)^{-1}t^{iI} \\
R^{i}&=&t^{oII}p^{o}r^{iII}p^{i}(1-w)^{-1}t^{iI}+r^{iI}
\end{eqnarray}
with $w=r^{oII}p^{o}r^{iII}p^{i}$ where $p^{i}$ and $p^{o}$ are the diagonal
propagation matrices for internal and external modes within the
curved region respectively. Similar equations can be obtained if the
incident modes come from the the other extreme of the bend
(Fig.\ \ref{fig5}a):
\begin{eqnarray}
T^{o}&=&t^{oII}p^{i}(1-w)^{-1}t^{oI} \\
R^{o}&=&t^{iII}p^{i}r^{oII}p^{o}(1-w)^{-1}t^{oI}+r^{oI}
\end{eqnarray}
It was also possible to undertake the problem as a matching of
three regions \cite{prevko} as in Sec.\ \ref{match} but numerical problems,
even for bends of a small angle, leads us to prefer the method above. Besides,
any number of bends can be joined in the same way
by combining the latter four matrices
through the diagonal propagation matrix $p$ of straight regions (see Sec.\
\ref{box}).

Fig.\ \ref{fig6}a shows results of the conductance for a
single bend (of $\theta$ angle equal to $\pi/2$) as a function of the Fermi
energy with the first subband occupied. The width of the wire is
$1\: l_{m}$ and four different inner
radii $R_{i}$ have been chosen. It can be seen
the way in which the softness of this inner radius affects,
crucially, the value of the conductance near the bottom of the subband but
it is not important for larger values of the Fermi energy. Fig.\ \ref{fig6}b
shows the behavior of the conductance for the
Fermi energy near the second subband. An additional characteristic is the
appearance of a dip when the second subband is about to enter. Again
the important parameter is the inner radius which, given in magnetic lengths,
shows how an increasing magnetic field reestablishes a perfect conductance
even near the bottom of the subbands. In this way we can expect perfect
conductance in these kind of bends for reasonably low values of the
magnetic field.

\section{CONCLUSIONS}
\label{conclusions}

The conclusions of the work can be summarized in the following points: a)
The matching technique
in the presence of a magnetic field is not a trivial problem and must be
faced carefully in order to obtain correct results for actual situations.
b) Although other techniques can be used for strips or straight waveguides,
that presented here allows to face in the same manner changes, either
in the potential or
in the geometry of the waveguide,
without changing the framework used. In addition to it,
experimental results can be reproduced and easily understood.

\section*{ACKNOWLEDGMENTS}

This work has been supported in part by the Comisi\'on Interministerial
de Ciencia y Tecnolog\'{\i}a of Spain under contract MAT 91 0201 and by
the Commission of the European Communities under contract SSC - CT 90 -
0020.

\appendix

\section{numerical solution of Schr\"odinger equation}

Schr\"odinger equation \ref{schr} can be
written in the following way (dropping all the labels from it):
\begin{equation}
-\phi"(x)+f(x,k)\phi(x)-\epsilon\phi(x)=0
\label{a1}
\end{equation}
where $f(x,k)=(x+k)^{2}+v(x)$.
If we discretized the $x$ variable into $N+1$ one-dimensional cells
(with $N$ of the order of $100$ in the actual calculations)
within the limits $x_{0}$ and $x_{N+1}$
so that $x_{i}=iq+x_{0}$ with $q=(x_{N+1}-x_{0})/(N+1)$ then,
$\phi(x_{i})=\phi_{i}$ and
the second derivative can be expressed (to the lowest order) as
\begin{equation}
\phi_{i}"=\alpha(\phi_{i-1}-2\phi_{i}+\phi_{i+1})
\end{equation}
with $\alpha=q^{-2}$. Eq.\ \ref{a1} becomes
\begin{equation}
-\alpha(\phi_{i-1}-2\phi_{i}+\phi_{i+1})+f(x,k)\phi_{i}-
\epsilon\phi_{i}=0
\end{equation}
With the following boundary conditions for the $\phi_{i}$ function
\begin{equation}
\left\{
\begin{array}{lll}
\phi_{0} & = & 0 \\
\phi_{N+1} & = & 0
\end{array}
\right.
\end{equation}
the Schr\"odinger equation \ref{schr} can be transformed into the
following homogeneous system of equations:
\begin{equation}
\left( \begin{array}{ccccc}
W_{1}(k) & -\alpha & 0 & ... &   \\
-\alpha  & W_{2}(k)& -\alpha & ... &   \\
    0    &         & \ddots  &     &  0\\
         &  ...    & -\alpha & W_{N-1}(k) & -\alpha \\
         &         &   0     & -\alpha & W_{N}(k)
\end{array} \right)
\left( \begin{array}{c}
\phi_{1} \\ \phi_{2} \\ \vdots \\ \phi_{N-1} \\ \phi_{N}
\end{array} \right) \; = \;
\left( \begin{array}{c}
0 \\ 0 \\ \vdots \\ 0 \\ 0
\end{array} \right)
\end{equation}
with $W_{i}(k)=[f(iq+x_{0}+k)-\epsilon]+2\alpha$.
Such equation system has a non-trivial solution if and only if
\begin{equation}
det\left( \begin{array}{ccccc}
W_{1}(k) & -\alpha & 0 & ... &   \\
-\alpha  & W_{2}(k)& -\alpha & ... &   \\
   0     &         & \ddots  &     &  0\\
         &  ...    & -\alpha & W_{N-1}(k) & -\alpha \\
         &         &   0     & -\alpha & W_{N}(k)
\end{array} \right) \; = \; D_{N}(k) \; = \; 0
\end{equation}
where $det$ denote the determinant which can be expressed by means
of a recurrence relation as
\begin{equation}
\begin{array}{lll}
D_{-1}(k) & = & 0 \\
D_{0}(k)  & = & 1 \\
D_{1}(k)  & = & W'_{N}(k) \\
          & \vdots &   \\
D_{i}(k)  & = & W'_{N-i+1}(k)D_{i-1}(k)-D_{i-2}(k) \\
          & \vdots &   \\
D_{N}(k)  & = & W'_{1}(k)D_{N-1}(k)-D_{N-2}(k) \; = \; F(k)
\end{array}
\end{equation}
with $W'_{i}=-W_{i}/\alpha$.
The complex roots $k_{n}$ of the so-defined
$F(k)$ function are the wave vectors we are
looking for. \\
\par
Once we know the wave vectors $k_{n}$ we can calculate the
corresponding $\phi_{i}(k_n)$ complex
functions to be used in the matching by means of
\begin{equation}
\phi_{i+1}(k_{n})=-\phi_{i-1}(k_{n}) +
[(P_{i}(k_{n})-\epsilon)/\alpha + 2]\phi_{i}(k_{n})
\end{equation}
where $P_{i}(k_{n})=f(iq+x_{0}+k_{n})$.

In order to check the functions obtained above we can resort to
standard Sturm-Liouville theory. Those functions must behave in a way
that the following expression must be obeyed (we return to the usual
continuos space $\phi_{i}(k_n) \equiv \phi_n(x)$):
\begin{equation}
(k_{i}-k^{*}_{j})\int_{x_{0}}^{x_{N+1}}
(k_{i}+k^{*}_{j}-2x)\phi_{j}^{*}(x)\phi_{i}(x) = 0
\end{equation}
\par
Eq.\ \ref{schrc} can be treated in exactly
the same way just by substituting $f(x,k)$ by
\begin{equation}
f(r,m)=-\frac{1}{4r^{2}}+\frac{1}{r^{2}}\left( \frac{r^{2}}{2}-m \right)^{2}
+ v(r)
\end{equation}
and the $\psi_{n}(r)$ functions must obey the expression
\begin{equation}
(m_{i}-m^{*}_{j})\int_{r_{0}}^{r_{N+1}}
((m_{i}+m^{*}_{j})/r^{2}-1)\psi_{j}^{*}(r)\psi_{i}(r) = 0
\end{equation}

\begin{figure}
\caption{Schematic view of a waveguide crossed by a barrier-like potential
showing the three regions $I,II,III$ involved in the matching problem.
The thin lines give an idea of
the trajectories followed by the edge states although they do not behave
adiabatically in this geometry any more. \label{fig1}}
\end{figure}
\begin{figure}
\caption{a) Dispersion relation of the lowest subbands in a waveguide
in the presence of a perpendicular magnetic field. It is also shown the
chosen energy above the three lowest subbands. b) All the
wave vectors found numerically for a given energy depicted in the $k$-complex
plane. The effect of either increasing or decreasing the magnetic field is
shown visually by the arrows pointing outwards or inwards
respectively. \label{fig2}}
\end{figure}
\begin{figure}
\caption{a) Conductance versus Fermi energy for a magnetic field of $1$ T.
The length $L$ of the box is $8\: l_{m}$ and the width
$W$ is $4\: l_{m}$. The height
of the barriers is $1.5\: \hbar\omega_{c}/2$
and the width of them $0.5\: l_{m}$.
The inset shows the semiclassical picture of trajectories followed
by the edge states in both
regions $A$ and $B$. b) Conductance for the same
box in the case of $1.56$ T. \label{fig3}}
\end{figure}
\begin{figure}
\caption{Conductance versus magnetic field for a Fermi energy
of 6 $\hbar\omega_{c}/2$ of a box of length $8\: l_{m}$, width $4\: l_{m}$,
height of the barriers $1.5\: \hbar\omega_{c}/2$
and width of them $0.5\: l_{m}$. \label{gb}}
\end{figure}
\begin{figure}
\caption{Evolution of the fifth state when increasing the magnetic field.
The picture shows how the density tends to stick to the walls forming
an edge state running all the way round the box. \label{fig4}}
\end{figure}
\begin{figure}
\caption{a) View of the total bend showing the two different possibilities
for incoming modes as explained in the text.
b) Schematic view of the matching between regions $I$ and $II$ of
different geometry. The arrows show the four different possibilities for
the incoming modes as explained in the text. \label{fig5}}
\end{figure}
\begin{figure}
\caption{a) Conductance of a
waveguide of $1\:l_m$ wide with a bend of angle $\pi/2$ for
different inner radii $R_i$ as a function of the Fermi energy. The first
subband is the only one occupied.
b) Conductance with the first and second subbands occupied
for different inner radii in the same case than a). \label{fig6}}
\end{figure}

\begin{references}

\bibitem{ssp441}C. W. J. Beenakker and H. van Houten, in {\em
Solid State Physics} vol. {\bf 44}, p. 1, edited
by D. Turnbull and H. Ehrenreich
(Academic, New York, 1991) and references therein.
\bibitem{zpb59385}See, e.g., A. Mac Kinnon, Z. Phys. B {\bf 59}, 385
(1985).
\bibitem{prb448017}T. Ando, Phys. Rev. B {\bf 44}, 8017 (1991).
\bibitem{ssc68715}G. Kirczenow, Solid State Comm. {\bf 68}, 715 (1988); G.
Kirczenow, Phys. Rev. B {\bf 39}, 10452 (1989); Y. Avishai and Y. B. Band,
Phys. Rev. B {\bf 40}, 12535 (1989).
\bibitem{prb403429}Y. Avishai and Y. B. Band,
Phys. Rev. B {\bf 40}, 3429 (1989).
\bibitem{prb4513725}J. J. Palacios and C. Tejedor, Phys. Rev. B {\bf 45},
13725 (1992).
\bibitem{prb4110354}P. F. Bagwell, Phys. Rev. B {\bf 41}, 10354 (1990)
\bibitem{prb439012} A. Kumar and P. F. Bagwell, Phys. Rev. B {\bf 43},
9012 (1991).
\bibitem{jpsj601679}H. Kasai, K. Mitsutake and A. Okiji, J. Phys. Soc. Jap.
{\bf 60}, 1679 (1991).
\bibitem{prb446315}H. Wu, D. W. L. Sprung, J. Martorell and S. Klarsfeld,
Phys. Rev. B {\bf 44}, 6351 (1991).
\bibitem{prb4111887}F. Sols and M. Macucci,
Phys. Rev. B {\bf 41}, 11887 (1990)
\bibitem{jap71515}D. W. L. Sprung, H. Wu and J. Martorell,
J. Appl. Phys. {\bf 71},
\bibitem{prb4511960} H. Wu, D. W. L. Sprung and J. Martorell,
Phys. Rev. B {\bf 45}, 11960 (1992).
\bibitem{prevko}K. Vacek, H. Kasai and A. Okiji, preprint.
\bibitem{prb4312082}J. A. Brum, Phys. Rev. B {\bf 43}, 12082 (1991).
\bibitem{prl622527}Y. Avishai and Y. B. Band, Phys. Rev. Lett. {\bf 62},
2527 (1989).
\bibitem{prb395476}R. L. Schult, D. G. Ravenhall and H. W. Wyld, Phys. Rev. B
{\bf 39}, 5476 (1989).
\bibitem{prb4112760}R. L. Schult, H. W. Wyld and D. G. Ravenhall, Phys. Rev. B
{\bf 41}, 12760 (1990).
\bibitem{prb448399}Y. Takagaki and D. K. Ferry, Phys. Rev. B {\bf 44}, 8399
(1991).
\bibitem{prange}{\em The Quantum Hall Effect}, edited by R. E.
Prange and S. M. Girvin (Springer-Verlag, New York, 1987).
\bibitem{jpsj592884}H. Yoshioka and Y. Nagaoka, J. Phys.
Soc. Jpn. {\bf 59}, 2884 (1990).
\bibitem{sslfs}S. E. Laux, D. J. Frank and F. Stern, Surf. Sci. {\bf 196},
101 (1988).
\bibitem{jpcm16291}R. J. Brown, C. G. Smith, M. Pepper, M. J. Kelly,
R. Newbury, H. Ahmed, D. G. Hasko, J. E. F. Frost, D. C. Peacock, D. A.
Ritchie and C. A. G. Jones, J. Phys. Condens. Matter {\bf 1}, 6291 (1989).
\bibitem{thkou}L. P. Kouwenhoven, PhD thesis, Delft University of Technology,
1992, (unpublished).
\bibitem{prb464026}D. B. Chklovskii, B. I. Shklovskii and L. I. Glazman,
Phys. Rev. B {\bf 46}, 4026 (1992).
\bibitem{prebpt}L. Brey, J. J. Palacios and C. Tejedor, preprint.
\bibitem{aclarar}Strictly speaking, the number of
wave vectors for these subbands should be also four
(with real parts very close to zero) but, due to the numerical
restrictions imposed by the discretization of space (see Appendix), only
two different ones can be found within the numerical accuracy required for our
calculations. This fact does not affect the results in the least.
\bibitem{prb3810507}L. Brey, G. Platero and C. Tejedor, Phys. Rev. B {\bf 38},
10507 (1988).
\bibitem{sct}{\em Single Charge Tunneling},
edited by H. Grabert and M. H. Devoret,
NATO ASI Series B (Plenum, New York, 1991).
\bibitem{prept}J. J. Palacios, L. Mart\'{\i}n-Moreno and C. Tejedor,
preprint.
\bibitem{prb441646}C. W. J. Beenakker, Phys. Rev. B {\bf 44}, 1646 (1991).
\bibitem{prep}C. J. B. Ford, P. J. Simpson, M. Pepper, D. Kern,
J. E. F. Frost, D. A. Ritchie and G. A. C. Jones, preprint.
\bibitem{prl622523}B. J. van Wees, L. P. Kouwenhoven, C. J. P. M. Harmans,
J. G. Williamson, C. E. Timmering, M. E. I. Broekaart, C. T. Foxon and
J. J. Harris, Phys. Rev. Lett. {\bf 62}, 2523 (1989).
\bibitem{jpcm13369}D. A. Wharam, M. Pepper, R. Newbury, H. Ahmed, D. G. Hasko,
D. C. Peacock, J. E. F. Frost, D. A. Ritchie and G. A. C. Jones, J. Phys.
Condens. Matter {\bf 1}, 3369 (1989).
\bibitem{prl691989}R. P. Taylor, A. S. Sachrajda, P. Zawadzki,
P. T. Coleridge and J. A. Adams, Phys. Rev. Lett. {\bf 69}, 1989 (1992).
\bibitem{prb467236}B. W. Alphenaar, A. A. M. Staring, H. van Houten,
M. A. A. Mabesoone, O. J. A. Buyk and C. T. Foxon, Phys. Rev. B {\bf 46},
7236 (1992).
\bibitem{premb}I. K. Marmorkos and C. W. J. Beenakker, preprint.
\end{references}
\end{document}